\documentstyle[11pt,epsf]{article}
\def\thefootnote{\fnsymbol{footnote}}
\newcommand{\rl}{\rule[-0.6cm]{0cm}{1.2cm}}

\newcommand{\sts}{\footnotesize}

\newcommand{\ba}{\begin{array}}
\newcommand{\ea}{\end{array}}

\newcommand{\ds}{\displaystyle}

\newcommand{\ri}{\right}
\newcommand{\lf}{\left}
\newcommand{\th}{\theta}

\newcommand{\eq}{\begin{equation}}
\newcommand{\en}{\end{equation}}
\newcommand{\bea}{\begin{eqnarray}}
\newcommand{\eea}{\end{eqnarray}}
\newcommand{\acc}{\\[3mm]}
\newcommand{\ZZ}{\hbox{{\rm Z{\hbox to 3pt{\hss\rm Z}}}}}
\newcommand{\Y}{\Upsilon}

\newcommand{\CM}{{\cal{M}}}
\newcommand{\CT}{{\cal{T}}}
\newcommand{\CC}{{\cal{C}}}

\newcommand{\virg}{~~,~~}
\newcommand{\pu}{~~.~~}

\newcommand{\Z}{{\cal Z}}

\newcommand{\CS}{S}

\newcommand{\NP}[1]{Nucl.\ Phys.\ {\bf #1}}
\newcommand{\PL}[1]{Phys.\ Lett.\ {\bf #1}}

\newcommand{\PRL}[1]{Phys.\ Rev.\ Lett.\ {\bf #1}}
\newcommand{\MPL}[1]{Mod.\ Phys.\ Lett.\ {\bf #1}}
\newcommand{\IJMP}[1]{Int.\ J.\ Mod.\ Phys.\ {\bf #1}}

\newcommand{\JMP}[1]{J. \ Math.\ Phys.\ {\bf #1}}
\newcommand{\JPS}[1]{J. \ Math.\ Phys.~Sci. \ {\bf #1}}
\newcommand{\PTP}[1]{ Prog.\ of \ Theo. \ Phys . {\bf #1}}
\newcommand{\PHY}[1]{ Physica {\bf #1}}
\newcommand{\JP}[1]{ J.\ Phys. \ {\bf #1}}

\newsavebox{\te}
\sbox{\te}{\begin{picture}(52,35)(0,0)
\put(10,20){\line(1,1){10}}
\put(10,20){\line(1,-2){10}}
\put(10,20){\line(5,-2){12.5}}
\put(30,20){\line(-1,1){10}}
\put(30,20){\line(-1,-2){10}}
\put(30,20){\line(-3,-2){7.5}}
\put(30,20){\circle*{.5}}
\multiput(20,30)(0,-0.5){60}{\circle*{.15}}
\put(20,30){\line(1,-6){2.5}}
\put(20,0){\line(1,6){2.5}}
\multiput(10,20)(1,0){20}{\line(1,0){.75}}
\put(20,-1){\makebox(0,0)[t]{{\protect  {\em e} }}}
\put(9,20){\makebox(0,0)[t]{{\protect   {\em c} }}}
\put(31,20){\makebox(0,0)[t]{{\protect  {\em d} }}}
\put(20,31.5){\makebox(0,0)[t]{{\protect  {\em a} }}}
\put(23.5,15){\makebox(0,0)[t]{{\protect  {\em b} }}}
\put(22.55,15.1){\circle*{.5}}
\put(20,0){\circle*{.5}}
\put(10,20){\circle*{.5}}
\put(20,30){\circle*{.5}}
\end{picture}}
\begin{document}
\begin{titlepage}
\vskip 0.5cm
\begin{flushright}
DFTT-27/95 \\
DTP-19/95 \\
May, 1995\\
{\tt hep-th/9505102}
\end{flushright}
\vskip0.5cm
\begin{center}
{\Large {\bf
Thermodynamic Bethe Ansatz}\\
{\bf
and}\\
{\bf
Threefold Triangulations}}\\
\end{center}
\vskip 0.6cm
\centerline{F. Gliozzi$^a$ and R.Tateo$^{a\,b}$}
\vskip 0.6cm
\centerline{\sl  $^a$ Dipartimento di Fisica
Teorica dell'Universit\`a di Torino}
\centerline{\sl Istituto Nazionale di Fisica Nucleare, Sezione di Torino
\footnote{e-mail: gliozzi@to.infn.it}}
\centerline{\sl via P.Giuria 1, I--10125 Torino, Italy}
\vskip .2 cm
\centerline{\sl $^b$ Department of Mathematics, University of
Durham\footnote{e-mail:
tateo@to.infn.it, roberto.tateo@Durham.ac.uk}}
\centerline{\sl South Road,  DH1 3LE  Durham, England} \vskip.6cm
\begin{abstract}
\vskip0.2cm
\noindent
In the Thermodynamic Bethe Ansatz  approach to 2D integrable,
 ADE-related quantum field theories one derives a set of
algebraic functional equations (a Y-system) which play a prominent
role. This set of equations is mapped into the problem of finding finite
triangulations of certain 3D manifolds. This mapping allows us to find a
general explanation of the periodicity of the Y-system. For the $A_N$
related theories and  more generally for 
the  various restrictions of the
fractionally-supersymmetric  sine-Gordon models,
  we find an 
explicit, surprisingly
simple solution of such functional equations in terms of a single
unknown function of the rapidity.
The recently-found dilogarithm functional
equations associated to the Y-system simply express the invariance of
the volume of a manifold for deformations of its triangulations.

\end{abstract}
\end{titlepage}

\setcounter{footnote}{0}
\def\thefootnote{\arabic{footnote}}
\section{ Introduction}
In the Thermodynamic Bethe Ansatz (TBA) approach \cite{yy}, the
renormalization group behaviour of a two-dimensional, integrable
quantum field theory is described by the ground state
energy $E(R)$ of the system on
an infinitely long cylinder of radius $R$. The equations known (or
conjectured) to give $E(R)$ are of the form
\eq
E(R)=-\frac1{2\pi}\sum_{a=0}^{a=N}\int_{-\infty}^\infty d\theta\;
\nu_a(\theta)\log\left(1+Y_a(\theta)\right)~~~,
\label{en}
\en
where the $Y_a(\theta)$ are $R$ dependent functions determined by a set
of coupled integral equations known as TBA equations; the $\nu_a(\th)$
are known functions describing the asymptotic behaviour of the
solutions: $Y_a(\theta)\to \nu_a(\theta)$ for $\theta\to\pm\infty$.
One of the main results of ref.~\cite{ys} was that any  solution
$\{Y_a(\theta)\}$ of the TBA equations satisfies a set of simple
functional algebraic equations, called the Y-system. Conversely it is easy
to show that a set of entire functions satisfying the Y-system with  a
suitable asymptotic behaviour is a solution of the TBA equations, thus
the Y-system encodes all the dynamical properties of the model.

In~\cite{ys}-\cite{rtv}  a large class of TBA systems
classified according to the ADET  Dynkin diagrams was proposed to
describe integrable perturbed coset theories.
The Y-systems  associated to all these models can  be written in terms
of an ordered pair $G\times H$ of $ADET$ Dynkin diagrams in the
following form
\eq
\hspace{ -3mm}
Y_a^b\lf(\th+ \imath { \pi \over \tilde{g}}\ri)
Y_a^b\lf(\th- \imath {\pi \over \tilde{g}}\ri)=
\prod_{c=1}^{r_G} \lf( 1+Y_c^b(\th) \ri)^{G_{a\,c}}\prod_{d=1}^{r_H} \lf( 1+{1
\over Y_a^d(\th)} \ri)^{-H_{b\,d}}
\hspace{ -3mm}
\label{yy}
\en
where $G_{a\,c}$ and $H_{b\,d}$ are the adjacency matrices of
the corresponding  $ADET$ Dynkin diagram, $\tilde{g}$ is the dual
Coxeter number of $G$,  $r_G$ and $r_H$
are the ranks of the corresponding  algebras.
These ADE related Y-systems do not exhaust at all the set of such
systems. For instance, recently a new class of Y-systems associated to
 sorts of bent $D_n$ diagrams describing the sine-Gordon models at
rational points has been described~\cite{t}.
The functional equations  (\ref{yy}) are  universal in the sense that
using different infra-red boundary conditions, they  describe
different theories or different regimes  of the same theory.

The $Y$ functions solving whatever Y-system with arbitrary boundary
conditions  have two general, intriguing properties.

One is that the $Y$'s are {\sl periodic} functions, as first pointed out
by Zamolodchikov ~\cite{ys}. In particular, for the set of Y-systems
described in Eq.(\ref{yy}), denoting with $\tilde{h}$ the
dual Coxeter number of $H$ one can verify by direct successive
substitutions or, in the high rank cases, by numerical computations
that~\cite{ys,rtv}
\eq
Y_a^b\lf(\th+ \imath  \pi{\tilde{h} +\tilde{g}
 \over \tilde{g}} \ri) = Y_{\bar a}^{\bar b}(\th ) \virg
\label{xx}
\en
where $\bar a$ and $\bar b$ denote the nodes of the Dynkin diagram
conjugate to $a$ and $b$.
Although this periodicity has many important consequences and is in
relation with  the conformal dimension of the perturbing operator in
the ultra-violet region, it has not been proven in a general way
until now.

The other intriguing property, recently pointed out in ref.\cite{gt},
is that the $Y$ functions solving Eq.(\ref{yy})
are the arguments of a new infinite family of functional equations for
the Rogers dilogarithm $L(x)$  of the from
\eq
\sum_{a=1}^{r_G} \sum_{b=1}^{r_H}
\sum_{n=0}^{\tilde{h}+\tilde{g}-1} L\left(\frac{\Y_a^b(n)}
{1+\Y_a^b(n)}\right) =
\frac{\pi^2}{6}r_G r_H \tilde{g} ~~~,
\label{su}
\en
where
\eq
\Y_a^b(n)=Y_a^b\left(\th+ \imath \frac{n\pi}{\tilde{g}}\right)~~n=0,1,2,
\dots
\en
They were verified by extensive numerical checks and proven only for
low rank cases. Thus they are (until now) at the same conjectural level
of the  periodicity discussed above and are strictly related to it: if
the $\Y$ variables were not periodic Eq.(\ref{su}) would be meaningless.

{}From a physical point of view,  these two properties play an essential
role in finding the deformed conformal theory associated to the
integrable model, indeed form these functional equations one can easily
derive the whole set of the dilogarithm  sum-rules yielding the
effective central charge of the UV fixed point ~\cite{al1}-\cite{nah},
while the period of the $Y$ function fixes, as already mentioned, the
conformal dimension of the perturbing field.

In this paper we describe a simple geometrical interpretation of these
two properties.
In particular, for the infinite family of models related to
$A_N$ and, more generally,
for the  new set of Y-systems~\cite{t} describing the
$\phi_{1,1,3}$-perturbations of  $SU(2)$-coset models with 
fractional supersymmetry,
we work out explicitly the solution  for arbitrary boundary conditions. 
For  the $A_N$ systems we also
find  a general proof
of the  dilogarithm functional identities.

\vskip .6 cm
\section{ Threefold Triangulations}

 Volume calculations in the three-dimensional hyperbolic space
involve the dilogarithm function of complex argument,  as is well known
in the mathematical literature since the time of Lobachevskij.
Similarly, that the Rogers dilogarithm of
{\sl real } argument is  related to volume calculation of 
other threefolds. In particular  it has been shown \cite{ds} that there is at 
least a manifold (a compactification of the universal covering
group of the projective SL(2,R) group) where
there are special or "ideal" tetrahedra whose four vertices
are parametrized by four real numbers . The
volume of the ideal tetrahedron of vertices $a,b,c,d$ 
depends only on the 
cross-ratio $x=(abcd)\equiv\frac{(a-c)(b-d)}{(b-c)(a-d)}$ and is given in 
terms of the Rogers dilogarithm by
\eq
vol(abcd)=L(x)-\frac{\pi^2}{6}~~~.
\label{vol}
\en
Notice that the dilogarithm $L(z)$ is a multivalued function, then the above
definition  has some ambiguity unless it is implemented by a consistent 
choice of the sheet of the Riemann surface associated to $L(z)$. We choose 
the sheet in which $\vert vol(abcd)\vert$ has its minimum value.

With an abuse of notation we shall use in the following the same symbol
$(abcd)$ to denote both the cross-ratio as well as an ideal
tetrahedron of vertices $a,b,c,d$ with the proper orientation, whenever
it does not generate confusion; when instead it is important to distinguish 
between a tetrahedron and its cross-ratio we shall use square brackets. 

We do not need to commit ourselves on the 
nature of the threefold $\CM$  where Eq.(\ref{vol}) holds, and we consider 
Eq.(\ref{vol}) as the {\sl definition} of volume of an ideal tetrahedron of a 
suitable $\CM$ threefold.

Using the properties of the cross-ratio and of $L(x)$ it is immediate to
see that this volume is independent of  even
permutations of the vertices, as it should be; notice also that there are
always two vertex renumberings, corresponding to two different orientations,
where the value of the cross-ratio $x$ belongs to the interval
$0\leq x\leq1$. Indeed if the ideal tetrahedron $T$ is associated to the
cross-ratio $(abcd)=x$ then the the thetrahedron $U$ of reversed orientation 
is associated to the cross-ratio $(acbd)=1-x$. Note that, according to 
Eq.(\ref{vol}), these two tetrahedra do not have in general the same volume, 
although the (negative) sign of the volume is the same. 
However, owing to the Euler relation
\eq
L(x)+L(1-x)=L(1)=\frac{\pi^2}6~~~,
\label{euler}
\en
we get $vol(abcd)+vol(acbd)=-\frac{\pi^2}6$ . This has a simple geometric
interpretation. First, let us define the boundary $\partial T$ of the
tetrahedron $T=[abcd]$ in the following standard way
\eq
\partial[abcd]=[abc]-[bcd]+[cda]-[dab]~~~,
\label{bound}
\en
where $[ijk]$ denotes an oriented triangle. An odd permutation of the vertices 
gives the reversed orientation, then, for instance $[ijk]=-[jik]$. It follows
that the chain $T+U=[abcd]+[acbd]$ satisfies the 3-cycle condition
\eq
\partial(U+T)=0~~~.
\en 
Thus we can view $U$ as  the 
complement $U=\CM\setminus T$ of $T$ and the 3-cycle $T+U$ is a triangulation of
the threefold $\CM$. We can define the volume of $\CM$ as the sum of the
volumes of the tetrahedra of the triangulation:
\eq
vol(T)+vol(U)=vol(abcd)+vol(acbd)=vol(\CM)~~~,
\label{volm}
\en
and $vol(\CM)= - \pi^2 /6$. The negative value of this volume is simply a
consequence of the sign chosen in the definition of volume in  
Eq.(\ref{vol}). 

An elementary analog of the above special triangulation can be found for
instance in the sphere $S_2$: a spherical triangle $t$ plus its complement 
$u=S_2\setminus t$ form a
triangulation of $S_2$ and the area of $S_2$ is simply the sum of the areas of
these two triangles. One can get other less trivial triangulations by
replacing one (or both) triangle(s), say $t$, with a chain $\sum_i t_i$ of 
triangles with the same orientation whose boundary is that of $t$: 
$\partial\sum_i t_i=\partial t$. We shall see that a similar procedure can be 
adopted for the ideal tetrahedra of $\CM$.

Consider indeed a domain formed by two adjacent tetrahedra 
$U_1\leftrightarrow(adbc)$
and $U_2 \leftrightarrow (becd)$ like in Fig.~\ref{fig1} .

\setlength{\unitlength}{2.5mm}
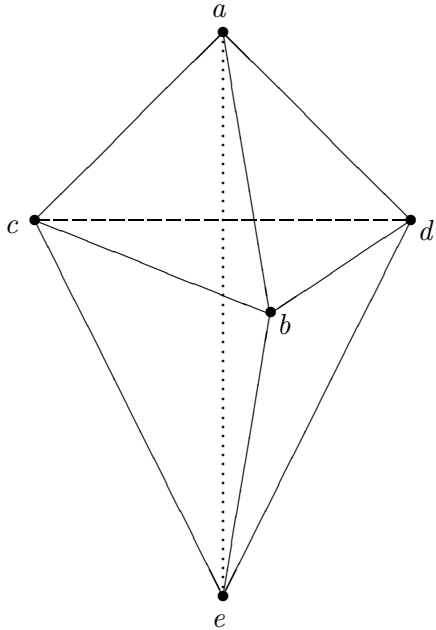
\begin{figure}[htbp]
\begin{center}
\begin{picture}(120,30)(0,0)
\put(5,5) {\usebox{\te}}
\put(3,0){\parbox{120mm}{\caption{\label{fig1} \protect
 {\sts Two adjacent tetrahedra can be divided into three. }}}}
\end{picture}
\end{center}
\end{figure}
It can also be considered as the
union of the three tetrahedra $T_3\leftrightarrow(cdae)\,,\,
T_4\leftrightarrow(deba) ~$ and
$T_5\leftrightarrow(eacb)~$:
\eq
U_1\cup U_2=T_3\cup T_4\cup T_5~~~,
\en
or, using   a more appropriate homology language, the chain
$C=U_1+U_2-T_3-T_4-T_5$ is a 3-cycle, i.e. the boundary of $U_1+U_2$ 
coincides with the boundary of $T_1+T_2+T_3$ so that $\partial C=0$.

Since the volume is additive with respect to division of the domain into
a finite number of pieces, we have
\eq
vol(U_1)+vol(U_2)=vol(T_3)+vol(T_4)+vol(T_5)~~~.
\label{dec}
\en
If the above tetrahedra are  ideal tetrahedra
of $\CM$ we get through Eq.(\ref{vol}) the five term Abel  functional 
identity ~\footnote{
The Abel and the Euler equations in the form~Eq.s (\ref{volm}) and (\ref{dec}) 
give us the rules  of passing from a tetrahedron
to its complement and  for  the decomposition of  two adjacent 
tetrahedra in to three. These  are the only two ingredients we need in 
 order to prove the more general $A_1 \times A_N$ (N$>$2) equations.} 
for the Rogers dilogarithm (see for instance ref.\cite{ds}) which 
coincides with the $A_1\times A_2$ case of Eq. (\ref{su}). 
However in this way
the hidden $\ZZ_5$ symmetry observed in ref.\cite{gt}, which played a
crucial  role in the construction of the new dilogarithm identities
does not have a clear geometric meaning. For a more symmetric approach, 
we combine Eq.(\ref{dec}) with Eq.(\ref{volm}), which is the geometric 
version of the Euler equation.
In particular, let us consider the complementary tetrahedra
$T_1=\CM\setminus U_1\leftrightarrow(abdc)$ and 
$T_2=\CM\setminus U_2\leftrightarrow(bced)$.
Clearly the chain $\CC=T_1+U_1+T_2+U_2$ is a 3-cycle, and Eq.(\ref{volm}) 
gives
\eq
vol(T_1)+vol(U_1)+vol(T_2)+vol(U_2)=2\,vol(\CM)~~~.
\en
Note that the volume of these four tetrahedra have the same sign, hence 
$\CC$ can be considered as a triangulation of  a $3D$ manifold
$\CM_2$ which covers  $\CM$ twice. 
Now, if  we replace the sum  $U_1+U_2$ with $T_3+T_4+T_5$ in $\CC$, we get a 
new triangulation of $\CM_2$
\eq
\CT=\sum_{i=1}^{5}T_i
\label{tri}
\en
which is {\sl regular}, in the sense that the group of the automorphisms
of the triangulation, denoted by $G_{\CT}$,  is transitive: in other
terms, given two arbitrary, distinct tetrahedra $T_i$ and $T_j$ there
is at least a permutation  of the tetrahedra preserving all the
adjacency relations
which carries $T_i$ into $T_j$.  Performing the cyclic permutation of
the five vertices $a\to b\to c\to d\to e\to a$ in the triangulation
$\CT$ yields $T_i\to T_{i+1}$ (the indices are taken modulo 5).
This shows that $\CT$ is regular and that $G_{\CT}\supset\ZZ_5~$.
Note also that, taking the vertices ordered as $a<b<c<d<e$,  the
cross-ratios associated to the 5 ideal tetrahedra belong to the
interval $[0,1]$ and the Abel identity (\ref{dec}) can be written in a more
symmetric way as 
\eq
\sum_{i=1}^5vol(T_i)=2\,vol(\CM)~~~.
\label{abel}
\en
\vskip .3 cm

It is now clear how generalize the above geometrical construction to
the other Y-systems: we conjecture that the $Y$ variables are
associated to the ideal tetrahedra of a triangulation $\CT$ of a
manifold  which is a multiple covering of $\CM$. This should imply
that writing the $Y$'s as cross-ratios of a suitable set of
points solves the Y-system equations. The $\ZZ_P$
symmetry of the Y-system ($P$ is the periodicity) is mapped into
the automorphism group $G_\CT$ of the triangulation. In the more
symmetric cases $G_\CT$ is transitive, yielding a regular triangulation.
Finally, the geometrical meaning behind the
Rogers dilogarithm functional identities of Eq.(\ref{su}) is simply that
the volume of a manifold is independent of its triangulations.

It is important to stress that, although we use a geometric language to 
express
our results, the whole set of  equations we write is independent of this
geometrical interpretation: all our derivations are simply algebraic 
consequences of the Euler and Abel identities given in Eq.s (\ref{volm}) 
and (\ref{abel}); nevertheless the geometric language  is extremely useful 
to guide our intuition.

\vskip.3cm
We worked out explicitly the construction described above for the
infinite family of the kind $A_1\times A_N$, and more generally for
the  new set of Y-systems~\cite{t} describing the
 various restrictions of the
fractionally-supersymmetric  sine-Gordon models. 
The solution is surprisingly 
simple.

We start by describing  the case $A_1\times A_3$ and then we will generalize to 
a generic $A_N$.

Consider an ordered set of six points $x_1<x_2<\dots<x_6$ of the real line which
will form the vertices of a regular triangulation of a suitable threefold. We
give to the indices of $x_i$  a cyclic order by putting
\eq
x_{i+6}=x_{i}~~~.
\label{cyc}
\en
Consider now the following three ideal tetrahedra
\eq
T_i\leftrightarrow(x_{i+2}\,x_{i+3}\,x_{i+1}\,x_i)\;\;i=1,2,3~~.
\en 
They have the same adjacency relations of the nodes of the $A_3$ diagram.
We now associate to each $T_i$ its complement 
\eq
U_i=\CM\setminus T_i
\leftrightarrow(x_{i+2}\,x_{i+1}\,x_{i+3}\,x_i)\;\;i=1,2, 3~~.
\label{t5}
\en
According to Eq.(\ref{volm}) we have 
\eq
\sum_{i=1,2,3}\left[vol(T_i)+vol(U_i)\right]=3vol(\CM)~~~.
\en
Then the 3-cycle $\sum_{i=1}^3[T_i+U_i]$ can be considered as a triangulation 
of a manifold $\CM_3$ covering $\CM$ three times. We now apply few times the 
five term relation (\ref{dec}) to pairs of adjacent tetrahedra
of the kind $U_i$, in order to obtain a regular triangulation. In the following
chain of relations the braces select the pair of such adjacent tetrahedra.

\eq
\ba{c}
\ds{\sum_{i=1}^3vol(U_i)=\{vol(3241)+vol(4352)\}+vol(5463)  = } \acc
\ds{vol(1254)+vol(2315)+\{vol(5143)+vol(5463)\} =} \acc
\ds{vol(1254)+vol(3416)+\{vol(2315)+vol(1365)\}+vol(6154) =} \acc
\ds{vol(1254)+vol(3416)+vol(5632)+}\acc
\ds{+vol(1265)+vol(2316)+vol(6154) } \acc
\ea
\en
Note that, as a consequence of the ordering $x_1<x_2<\cdots$ of the vertices, 
all the cross-ratios belong to the interval $[0,1]$ and all the volumes of the 
tetrahedra listed in the above identities have the same sign.
If we add to the three $T_i$'s of Eq.(\ref{t5}) these six tetrahedra which 
replace the $U_i$'s  we get a {\sl regular} triangulation $\CC$ of $\CM_3$.
Indeed these nine ideal tetrahedra are associated
to the cross-ratios of the form $(x_{j}\,x_{j+1}\,x_{i+1}\,x_i)
\leftrightarrow T_{ij}~$ $(T_{i,j} = T_{j,i})$, 
where $i$ and $j$ are two non-consecutive, cyclic indices 
($i+6=i\;,\;j+6=j$).
The initial $T_i$ tetrahedra  correspond, in this notation, to $T_{i\,
i+2}$. In order to study the adjacency relations of these nine tetrahedra it is
useful to draw a graph like in Fig.2 
\setlength{\unitlength}{1.mm}
\begin{figure}
\begin{center}
\vspace{-1.cm}
\leavevmode
\epsfysize=7cm \epsfbox{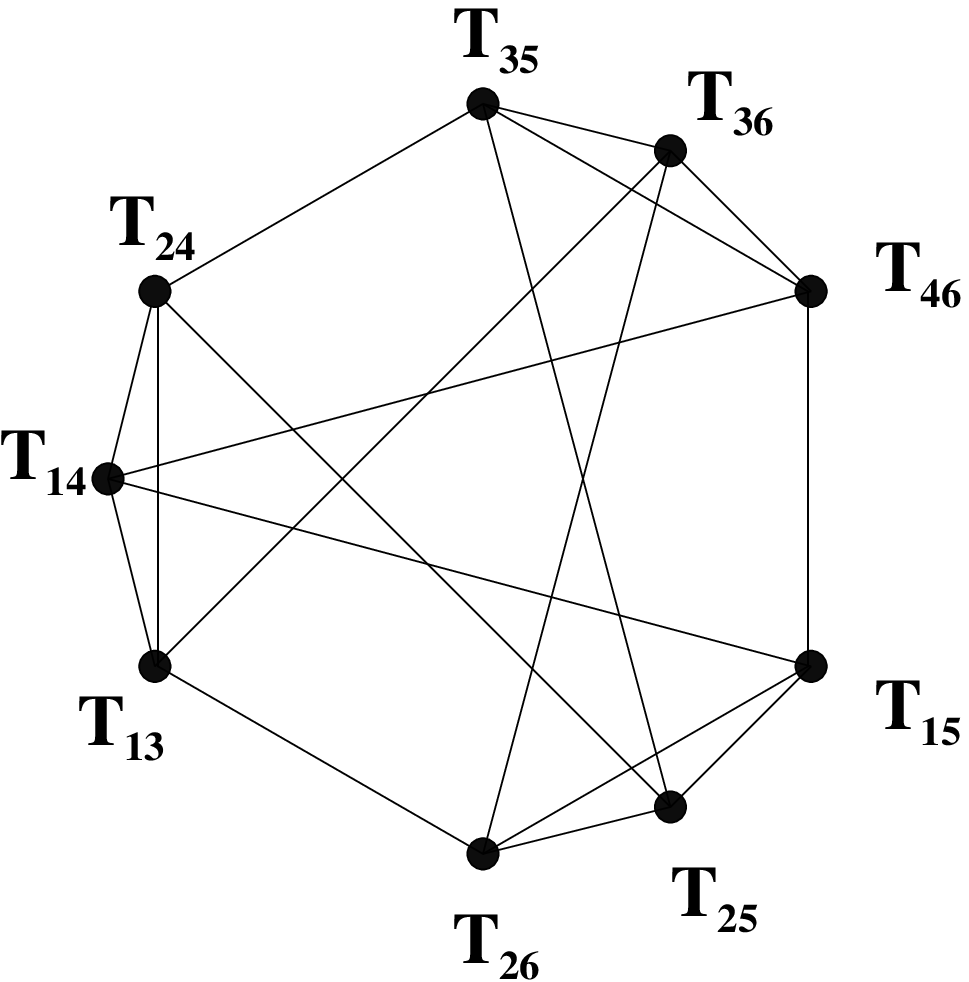}
\put(-30,-10){\makebox(0,0)[t]{{Figure~2: Adjacency relations for the 
$A_3$ triangulation.}}} 
\end{center}
\end{figure}
where the nodes are associated to  the tetrahedra $T_{ij}$ and adjacent nodes 
represent adjacent tetrahedra. It is easy to verify that this graph, 
besides the obvious 
$\ZZ_3$ cyclic symmetry $T_{ij}\to T_{i+2,j+2}~$, fulfils another more hidden
$\ZZ_3$ symmetry generated by  $(T_{13}T_{14}T_{24})(T_{35}T_{36}T_{46})
(T_{15}T_{25}T_{26})~$, where $(xyz)$ denotes the cyclic permutation 
$x\to y\to z\to x$. Combining
 these two symmetries it is evident that each $T_{ij}$ can be mapped in any 
other $T_{i'j'}$ as required in a regular triangulation.

The generalization of the above construction to a generic $A_N$ is 
straightforward: the starting object is an ordered set of $N+3$ points
$x_1<x_2<\dots<x_{N+3}$  which will form the vertices of a regular
triangulation. Being the cross-ratio invariant under the projective group,
one can perform a projective transformation carrying three
arbitrary distinct points of this set into $0,1, \infty$. As a
consequence this object depends on $N$ real parameters. 
\noindent
Consider now the following set of $N$ ideal tetra\-hedra
$T_i\leftrightarrow(x_{i+2}\,x_{i+3}\,x_{i+1}\,x_i)\;$ $\;i=1,2\dots N$
having the same adjacency relations of the nodes of the $A_N$ diagram.
These are the seeds of the wanted triangulation. Indeed
these with their complements $U_i=\CM\setminus T_i$
form a 3-cycle where, using again Eq.(\ref{volm}), we get
$\sum_{i=1}^{N}[vol(T_i)+vol(U_i)]=N\,vol(\CM)$, then it can be
thought as a triangulation of a manifold $\CM_N$ covering $\CM$ $N$
times. One can try again to transform this triangulation into a regular one
by using the decomposition of Eq.(\ref{dec}) in suitable places. Actually
 a good strategy is to eliminate all the tetrahedra of
kind $U_i$, because pair of complementary tetrahedra 
as $\{U_i,T_i\}$ are easily seen to be obstructions to the transitivity
of the automorphisms. For the general $A_N$ case one ends up
with a chain $\CC$ of $N(N+3)/2$ ideal tetrahedra $T_{ij} $ associated
to the cross-ratio $(x_{j}\,x_{j+1}\,x_{i+1}\,x_i)~$, labelled by two
non-consecutive indices, with $1<i<i+1<j\leq N+3$ or $2<j<N+3$ if $i=1~$,
where the indices are understood to be cyclically ordered by defining
$x_{i+N+3}=x_{i}$. Notice that all these cross-ratios belong to the
interval $[0,1]$ as a consequence of the ordering of the vertices.

Now it is easy to show directly that $\CC$ is a regular triangulation of
$\CM_N$: Note that the  chain $\CC$
satisfies the 3-cycle condition because, as is immediate to verify using 
Eq.(\ref{bound}), each face of the $T_{ij}$'s is shared by two tetrahedra 
and appear with the two opposite orientations, then $\partial\CC=0$. 
If we subtract from $\CC$ the chain of seed tetrahedra, we
get a chain $\CC'=\CC-\sum_{i=1}^{ N}T_i$ with the same vertices of
of $\CC$. Now  $\CC'$ is no longer a cycle, but has a boundary which
coincides with the boundary of $\sum_{i=1}^{N}U_i$. It follows that
$\CC'$ is a triangulation of $\sum_{i=1}^{N}U_i$ and
$\CC$ is a triangulation of $\CM$. Looking at the adjacency relations
it is not difficult to see  that it is regular.

As a first consequence, the volume of $\CM_N$ can be written as
\eq
\sum_{T_{ij}\in\CC} \,vol\left(T_{ij}\right)=N\,vol(\CM)~~~,
\label{vola}
\en
which is nothing but the Rogers dilogarithm identity for the
$A_N\times A_1$ system in a slightly disguised form.
Indeed, rewriting  Eq.(\ref{yy}) for this family of diagrams as
\eq
Y_a\lf(\th+ \imath { \pi \over {N+1}}\ri)
Y_a\lf(\th- \imath {\pi \over {N+1}}\ri)=
\prod_{c=1}^{N} \lf( 1+Y_c(\th) \ri)^{G_{a\,c}}~~~,
\label{yyan}
\en
the identity (\ref{su})  in its~\cite{gt} reduced form becomes
\eq
\sum_{a=1}^N\sum_{n=0}^{N+2}\delta(a+n)L\left(\frac{\Y_a(n)}{1+\Y_a(n)}\right)
=\frac{\pi^2}{6}N(N+1)
\label{ida}
\en
where the projector $\delta(j)=(1+(-1)^j)/2$ constrains the double sum
to the subset in which $a+n$ is an even number and
\eq
\Y_a(n)=Y_a\left(\th+ \imath \frac{n\pi}{N+1}\right)~~n=0,1,2,
\dots N+2~~~.
\label{yan}
\en
Then we have to map the two indices $i,j$ labelling the ideal
tetrahedra into the two indices $a=1,\dots N$ and $n=1,\dots N+3$.
It is convenient to choose
\eq
  i=\left[\frac{n-a}{2}\right]-1~~~,~~~j=\left[\frac{n+a}{2}\right]
{}~~~,
\label{map}
\en
where $[x]$ denotes the integer part, modulo $N+3$. An obvious
permutation of the indices in the cross-ratio yields
\eq
\Y_a(n)=-(x_{j}\,x_i\,x_{i+1}\,x_{j+1}\,)~~~.
\label{one}
\en
With this choice Eq.(\ref{vola}) is transformed, using Eq.(\ref{vol}),
into the Eq.(\ref{ida}) and the Y-system (\ref{yyan}) is identically
fulfilled as a consequence of the following trivial identity among cross
-ratios involving six different points
\eq
(ebcf)(dabe)=(ebaf)(dcbe)~~~.
\en
In particular, for variables associated to the two end nodes of the
$A_N$ diagram the number of different points in the above identity is
reduced to five and one  of the two cross-ratios in the r.h.s is equal to
1 as it should.

An important feature of the map (\ref{map}) is that the indices $a$ and
$n$, which in the Y-system have a completely different role (the former
labels the nodes of $A_N$ the latter is the recursion index) here appear
summed together to form the indices labelling the vertices $x_i,x_j$
of the triangulation. Since $n$ in  Eq.(\ref{yan}) enters in the
imaginary part of the rapidity $\th$, these vertices can be thought as
the different values of a {\sl single} periodic function $x(\th)$ for
different values of the imaginary part of the argument. Thus,
combining Eq.s(\ref{yan}), (\ref{one}) and (\ref{map}) we may put
\eq
x_j=x \lf(\th+\imath \pi \frac{j}{N+1} \ri)
\label{xth}
\en
with the periodicity condition
\eq
x \lf(\th +\imath\pi\frac{N+3}{N+1}\ri)=x(\th)~~~.
\label{xper}
\en

In conclusion, inserting Eq.(\ref{xth}) in Eq.(\ref{one}) we get a solution 
of the Y-system functional equations for the $A_1\times A_N$ case. It is
immediate also to see that such a solution is the most general one, because 
it is known that this Y-system can be considered from the algebraic point of 
view as a recursion relation depending on $N$ free parameters, which is just 
the number of free parameters in our construction. According to Eq.(\ref{xper}) 
the general solution of the Y-system is periodic, so we get, as a by-product, 
a simple proof of the periodicity conjecture.

Finally, notice that the whole set of $Y$-functions is parametrized through
Eq.(\ref{one}) by a single  function $x(\th)$ which becomes
the only unknown of  the TBA equations.  It is suggestive that this
result  goes in the  direction 
of~\cite{ddv} where an alternative 
version for the XXZ vacuum energy, depending only from one unknown function, 
has been derived.

\section{ Triangulations for minimal models }

Among the integrable theories,  the sine-Gordon model  and its
quantum-reduced versions (RSG) corresponding to  the $\phi_{13}$-thermal
perturbations of the minimal $c < 1 $ conformal field theories, are
certainly  the more  studied in
the literature. The thermodynamic equations for  the vacuum energy  of  these
models has been determined  in the repulsive phase~\cite{tak}
as well in the attractive~\cite{t} one. In particular in~\cite{t}
 the algebraic RSG-functional equations, obtained using
the S-matrix of  the sine-Gordon model
at arbitrary rational coupling constant, have been presented.
It is important to stress~\cite{fi,tat,t}  that  these  set of 
equations describe not 
only the thermal perturbation of the $c<1$ minimal models but indeed  the
whole set of reduced models associated to the    
fractionally-supersymmetric  sine-Gordon theories~\cite{bl}.  
In the following we will report the general solutions for  the
Y-systems, proposed in that 
paper, and the reader should consult directly that work  for a correct
interpretation of our results.
\noindent
Let us remember that  the RSG Y-systems  are uniquely  defined
by the simple-continued fractions representation  for the "dressed ''
coupling
constant $\xi$ of the SG model
\eq
\xi={p \over q-p}=\hat{\xi}(n_1,n_2, \dots , n_F ):=
 { { 1 \over n_1 + {1
\over n_2 + \dots {1 \over n_F-1}}}} \virg
\label{con}
\en
and that the TBA equation can be written in term of a set of unknown 
functions $Y_k$ with  $k=1,2,\dots,n_T-3$ and 
$n_T= \sum_{i=1}^{F} n_i$.
One of the  main differences
between the RSG system and the $ADE$ is that the  shifts involved in the
Y-equations
are now index-dependent and in general they are linear combinations of
\eq
s_1=\imath \pi { \xi_1  \over 2 }
 ~~, ~~
s_2=\imath \pi {\xi_1 \xi_2 \over 2} ~~,~~ \dots ~~,~~ s_F=\imath \pi {
\xi_1 \xi_2 \dots \xi_F \over 2} \equiv {\imath \pi \over 2 q-2 p} ~~, \en
with
\eq
\xi_a=\hat{\xi}(n_a,n_{a+1}, \dots , n_F)
\pu
\en
Defining   the matrix
\eq
\hspace{ -3. mm}
c_{j,k}=  \left\{ \begin{array}{ll}
 0  & \mbox{for $j \ne k \pm 1$ } \\
 (-1)^{a-1} & \mbox{for  $ j  = k \pm 1$  , } 
 \sum_{i=1}^{a-1} n_i <
j,k \le 1+\sum_{i=1}^{a} n_i  \\
\end{array} \right.
\en
$j,k=1,2, \dots, n_T-3$.
The Y's are the solutions of the following system of coupled equations
($\tilde{c}_k=c_{k,k+1}$).
\[
Y_k \lf(\th+\CS_k\ri)~Y_k \lf(\th-\CS_k\ri)~=~(1+Y_{k-1}(\th)^{c_{k,k-1}}
)^{c_{k,k-1}}
~(1+Y_{k+n_{a+1}+1}(\th)^{\tilde{c}_{k}
})^{\tilde{c}_{k}}  
\]
\eq
\prod_{j=k+1}^{k+n_{a+1}}
(1+Y_{j}(\th + (k+n_{a+1}-j)
\CS_j+\CS_{k+n_{a+1}+1})^{\tilde{c}_{k}})^{\tilde{c}_{k}}
\label{y0}
\en
\[
\prod_{j=k+1}^{k+n_{a+1}}
(1+Y_{j}(\th - (k+n_{a+1}-j)
\CS_j-\CS_{k+n_{a+1}+1})^{\tilde{c}_{k}})^{\tilde{c}_{k}} \virg
\]
for  $k=\sum_{i=1}^a n_i$ and $ a < F-1$, with
\eq 
\CS_k= s_a \hspace{1. cm} \sum_{i=1}^{a-1}
n_i < k \le \sum_{i=1}^{a} n_i
\virg
\en

\[
Y_k\lf(\th+\CS_k\ri)~Y_k \lf(\th-\CS_k\ri)~=~
(1+Y_{k-1}(\th)^{c_{k,k-1} })^{c_{k,k-1}}~
(1+Y_{n_T-3}(\th)^{-\tilde{c}_k })^{-\tilde{c}_k}
\] 
\eq
\prod_{j=k+1}^{n_T-3}
(1+Y_{j}(\th+(n_T-1-j) \CS_j)^{\tilde{c}_k})^{\tilde{c}_k}
\en
\[
\prod_{j=k+1}^{n_T-3}   
(1+Y_{j}(\th-(n_T-1-j) \CS_j)^{\tilde{c}_k})^{\tilde{c}_k}
\]
for  $k=n_T-n_F$, and finally 
\eq
Y_k\lf(\th+\CS_k \ri)~Y_k\lf(\th-\CS_k \ri)~=~ \prod_{j}
(1+Y_j^{c_{j,k}}(\th))^{c_{j,k}}  \virg
\label{y1}
\en
for all the  other values of $k$.
For our purposes it is convenient  to introduce   a new set of 
positive integers
$\{ \tilde{n}_a \} $  defined in terms of the $\{n_a \}$ in 
Eq.~(\ref{con}) as
$\tilde{n}_1 =n_1+1$ ,  $\tilde{n}_F =n_F-1$ ,  $\tilde{n}_a = n_a$ ,
$ a=2,\dots F-1$
and the  integer   shifts  $\tilde{S}_a= 2(q-p) s_a / \imath \pi$.
With the choice $\tilde{S}_0=q$ they  satisfy the relation
\eq
\tilde{S}_{a-1} = \tilde{n}_a \tilde{S}_a + \tilde{S}_{a+1} \pu
\en
Let us also   define a  new set of functions labelled by two indices
$a$ and $i$
\eq
\rl \Z^a_i(n)^{{(-1)^{a+1}}} = \Y_k(n) \equiv
Y_k \lf( \th + \imath \pi { n \over 2 (q-p) } \ri),
\label{yss}
\en
with  $n=0,1,\dots, 2q-1$.
The  indices   $ a$ and  $i$ are defined   through  the relations 
\eq
k= \sum_{j=1}^{a-1} \tilde{n}_j + i  \virg
\en
\eq
\hspace{ -3. mm}
 \begin{array}{ll}
1 \le i \le \tilde{n}_a-1 
  & \mbox{ for $a=1$ } \\
0 \le i \le \tilde{n}_a-1
  & \mbox{ for $a=2,\dots,F-1$} \\
0 \le i \le \tilde{n}_a-3  
 & \mbox{ for $a=F$ } \\
\end{array}
\pu
\en
Our starting object is  now a ordered set of $q$ points
$x_1 < x_2 < \dots <x_q$, as in the  $A_N$ case
three of these can be fixed to be
 0,1,$\infty$ for the cross-ratio invariance under projective 
transformations. Using the experience on the $A_N\times A_1$ systems we can 
again try to solve explicitly the Y-system of these RSG 
models by expressing  quantities~(\ref{yss}) as suitable
cross-ratios of the above $q$ points. Actually it is straightforward to verify 
by direct substitution that the general solution of 
Eqs. (\ref{y0}-\ref{yss})  is
\eq
\Z^a_i(n) = - ( x_b~ x_c ~ x_d ~ x_e )
\label{zy}
\en
with $x_{i+q}=x_i$ and  
\eq
\ba{c}
\ds{ \rl
b=\lf[{ n - \tilde{S}_{a-1} + i \tilde{S}_{a} \over 2 } \ri]    \virg
c= \lf[{n + \tilde{S}_{a-1} - (i+2) \tilde{S}_{a} \over 2} \ri] \virg
}\acc
\ds{\rl
d=\lf[{n + \tilde{S}_{a-1} - i \tilde{S}_{a} \over 2} \ri] \virg
e=\lf[{n - \tilde{S}_{a-1} + (i+2) \tilde{S}_{a} \over 2 } \ri] \virg
}
\ea
\label{la}
\en
where $[x]$ denotes the integer part, modulo $q$.
$\Z^1_0$ is identically equal to  0 and  for  $\xi=1/(N+2)$
Eq.~(\ref{la}), up to an irrelevant common shift $q$ ,  
reduces to Eq.~(\ref{one}). 

As a simple, illustrative example, let us consider the theory at $\xi=3/4$. 
We have
$\tilde{n}_1=2$ , $\tilde{n}_2=3 $ , $\tilde{S}_1=3$ , 
$\tilde{S}_2=1$ , 
from Eq.s~(\ref{zy},\ref{la}) we find the fourteen tetrahedra of  this 
triangulation 
\eq
\hspace{ -3. mm}
T_n=  \left\{ \begin{array}{ll}
 ( [ (n-4)/2 ]~[(n-2)/2]~[(n+4)/2]~[(n+2)/2] )  & \mbox{for n even} \\
 ( [(n-3)/2]~[(n-1)/2]~[(n+1)/2]~[(n+3)/2] ) & \mbox{for n odd} \\
\end{array} \right.
\en
and $n=0,1, \dots 13$.
Drawing the adjacence diagram (see Fig.3), it is easy to see that, although 
the diagram has an evident $\ZZ_7$ symmetry, the triangulation, due to the  
asymmetry under the exchange $T_{n=odd} \leftrightarrow T_{n=even}$ , is not 
regular. 
\setlength{\unitlength}{1.mm}
\begin{figure}
\begin{center}
\vspace{-1.cm}
\leavevmode
\epsfysize=7.5cm \epsfbox{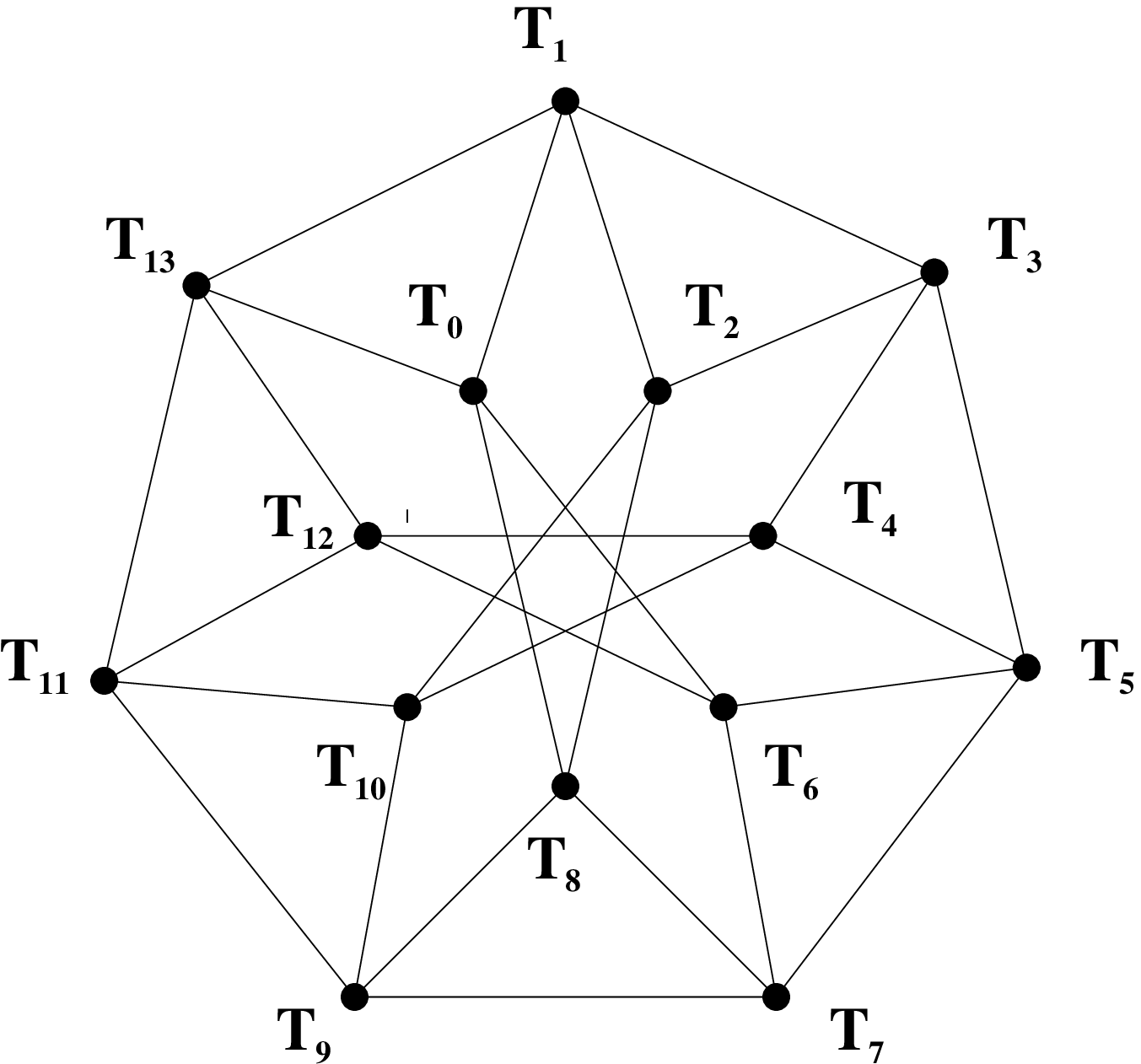}
\put(-30,-10){\makebox(0,0)[t]{{Figure~3: Adjacency relations for the 
$M_{3,7}$ triangulation.}}} 
\end{center}
\end{figure}

\section{  Conclusions}
The dilogarithm functions play an important role in various branches 
of mathematics and physics~\cite{le}-\cite{ki}. In particular, all the 
properties studied in this paper for the Rogers dilogarithm have an 
($m=2$) Bloch-Wigner counterpart~\cite{za}. The Bloch-Wigner   
function is used for volume calculation on hyperbolic manifold, and  
obviously  some of the geometrical concepts introduced  in this article
have been  already used in different contexts by various 
authors. However we only use the geometry as guide to our proofs, which 
are ultimately purely algebraic. We would also  like to stress 
that,  although we learned that the  $\ZZ_5$ symmetry in the Abel 
equation  noticed in~\cite{gt} was already known by the author 
of~\cite{za1}, to our knowledge the main 
results presented in this paper and in~\cite{gt,t} are new to the 
mathematics as well to the physics literature.\\

It was  known for long time that a three-dimensional viewpoint about
2D conformal field theories yields a better unifying understanding of
them: many different aspects of rational conformal field theories have
emerged as natural consequences of  the topology of three
dimensional manifolds. Our paper indicates that it is possible to
enlarge  this point of view also to  the perturbed conformal field
theories: different properties of the renormalization group
evolution of a 2D integrable theory perturbed with a relevant operator,
which are encoded in apparently mysterious properties of the Y-systems,
like their link with the Dynkin diagrams of simply-laced Lie algebras,
 the $\ZZ_P$ symmetry of the recursion relations, the  periodicity of
the $Y$ variables and the functional identities of the dilogarithm,
find again a unifying three-dimensional viewpoint through the ideal
triangulations of suitable threefolds.

\vskip .6 cm
{\bf Acknowledgements} --
R.T. would like to thank E.Corrigan , P.Dorey and R. de Jeu for useful 
discussions and the
Mathematics Department of Durham University for the kind hospitality.

\end{document}